\documentclass[%
preprint,
 amsmath,amssymb,
 aps,
]{revtex4-2}

\usepackage{graphicx}% Include figure files
\usepackage{dcolumn}% Align table columns on decimal point
\usepackage{bm}% bold math
\usepackage{appendix}
\usepackage[utf8]{inputenc}
\usepackage{epstopdf, epsfig}
\usepackage{subcaption}
\usepackage{float}
\usepackage{gensymb} % Required for degree symbols
\usepackage{hyperref}
\hypersetup{hypertexnames=false} 
\hypersetup{
    colorlinks=true,
    linkcolor=blue,
    citecolor=blue,
    filecolor=blue,      
    urlcolor=blue,
    linktocpage=True
    }
\usepackage[font=footnotesize]{caption}
\usepackage{siunitx}

\usepackage{tabularray}
\NewDocumentEnvironment{fancytblr}{+b}{
\begin{tblr}{
  row{1-Z} = {font=\footnotesize},
  hline{1,Z} = {0.08em},  
  hline{3} = {0.05em},
  cells = {c},
  }
#1
\end{tblr}
}{}

\newcommand{\hotspot}{$\mathbf{x_+}$}

\newcommand{\Bth}{$B_\text{th}$}
\newcommand{\thetamax}{$\Theta$} 
\newcommand{\thetathresh}{$\Theta_\text{th}$} 
\newcommand{\bond}{$\text{Bo}$}

\begin{document}

\title{Geometric and kinematic indicators of breaking inception in surface gravity waves}

\author{Daniel G. Boettger}
\author{Shane R.  Keating}
\author{Michael L. Banner}
\author{Russel P. Morison}
\author{Xavier Barth\'el\'emy}
\affiliation{School of Mathematics and Statistics, University of New South Wales,
Sydney, Australia}

\date{\today}

\begin{abstract}
The process of breaking in surface gravity waves can be characterized by two distinct stages. Breaking onset, defined as the first visible surface manifestation of breaking, is preceded by breaking inception, which is characterized by the initiation of an irreversible process within the crest that leads inevitably to breaking. Breaking inception diagnostics formulated using the local energetic, kinematic and geometric properties of the wave crest have recently been proposed that appear to provide generic parametric threshold estimates and may facilitate the effect of wave breaking on larger scale processes to be parameterized. In a recent numerical study [McAllister \textit{et al}, \textit{J. Fluid Mech.} 974, A14 (2023)], a breaking inception diagnostic was proposed based on a threshold value for the maximum local interface angle \thetamax{}. We extend these findings to include surface tension effects, which are an inescapable feature of ocean surface waves and are known to have a non-negligible effect on wave geometry. We conduct a suite of numerical breaking wave packet simulations, which incorporate surface tension, to show that \thetathresh{}$=60\degree$ is a robust threshold value for gravity waves of varying packet size, water depth and wind speed forcing. This value is twice the magnitude of that for waves in the absence of surface tension. We explore this result in the context of the kinematic inception parameter $B$ [Barthelemy \textit{et al}, \textit{J. Fluid Mech.} 841, 463 (2018)] and show that the kinematic and geometric methods are in fact equivalent and incorporate the same aspects of the underlying physical processes leading to wave breaking.
\end{abstract}

\maketitle
\section{Introduction}\label{s:intro}

The interaction between the atmosphere and ocean is a complex process of fundamental importance for the Earth system. The rate of momentum, energy, heat and gas fluxes are not only dependent on the dynamical properties of air and water, but are further modulated by the characteristics of the surface wave field. Breaking waves in particular drive enhanced fluxes into the ocean boundary layer \cite{dasaro2014} and quantifying the distribution of waves that break versus those that do not is critical in accurately characterizing this complex region. Wave breaking can be conceptually understood as a two-stage process. The first stage, breaking inception, describes the initiation of an irreversible process within the wave crest. This leads to breaking onset, defined as the first surface manifestation of breaking \cite{derakhti2020}. Wave breaking inception is a useful concept as it suggests the possibility of identifying a breaking wave in the earlier stages of its evolution when the process is more easily parameterized and represented in larger scale models.  

Efforts to characterize breaking inception are typically based upon diagnostic parameters that are linked to the  dynamic, kinematic or geometric properties of the wave \cite{perlin2013}. It is generally accepted that a complete understanding of the breaking process must be based upon fundamental fluid dynamics. However, the difficulty in observing the dynamical processes within an evolving wave has been a considerable impediment to progress in this field. Numerical experiments, which permit the diagnosis of all dynamical fields within the wave, have been successfully used to examine breaking inception \cite{barthelemy2018, derakhti2018, derakhti2020}. Recently, \citet{boettger2023} utilized a high-resolution numerical modeling framework to examine wave breaking and identified an indicator of breaking inception based upon the local kinetic energy balance.  This energetic indicator for breaking inception was shown to be robust for waves in deep and intermediate water depths, with varying wave packet sizes, and, in a subsequent numerical investigation, under varying wind forcing speeds \cite{boettger2024}. 

Numerical simulations allow the dynamics of the wave to be studied in detail. However, the results must be validated in field or laboratory experiments and this can be technically challenging for dynamical breaking inception parameters. Consequently, more easily studied kinematic and geometric approaches are widespread, despite the concession that they can only act as a proxy to the underlying dynamical causes of breaking. 

Kinematic breaking onset criteria are based on the hypothesis that a surface gravity wave will break if the horizontal fluid particle speed $\| \mathbf{u} \|$ exceeds the speed $\| \mathbf{c} \|$ of the wave. At first, this might be considered a reasonable conceptual model of breaking onset. However, a range of threshold values has been reported in the literature, many of them less than unity.  Much of this disparity may be attributed to the difficulty in robustly defining and measuring $\| \mathbf{c} \|$, which for unsteadily evolving waves typical of real-world conditions may vary between 0.7 -- 1.1 times the wave phase speed as it evolves from growth to decay \cite{fedele2020}. In fact, the kinematic ratio has been shown in laboratory experiments to vary by up to \mbox{55 \%} depending on the definition of $\| \mathbf{c} \|$, with values of $\| \mathbf{u} \|/\| \mathbf{c} \|$ at breaking onset as low as 0.68 \cite{stansell2002}.  

Without a dynamical underpinning on which to construct $\| \mathbf{u} \|/\| \mathbf{c} \|$, the variability in reported kinematic threshold values could be expected. In an effort to resolve this deficiency, \textcite{barthelemy2018} derived the kinematic ratio from fundamental fluid dynamic arguments. They began with the hypothesis that breaking onset is the result of a wave being unable to accommodate a local wave energy flux which exceeds that in the corresponding non-breaking case. The diagnostic parameter $B$ therefore represents the normalized ratio of the local energy flux to the local energy density. At the air-water interface, where the work against air pressure is negligible, this reduces to
\begin{equation}\label{eq:B}
    B = \frac{\| \mathbf{u} \|}{\| \mathbf{c} \|}, 
\end{equation}
where $\| \mathbf{c} \|$ is taken as a local measure of the unsteadily evolving speed of the crest tip. \citet{barthelemy2018} found that a threshold value $B_{th} = 0.855 \pm 0.05$ exists that separates breaking and non-breaking waves such that if $B_{th}$ is exceeded the crest will always evolve to break. Subsequent laboratory and numerical studies \cite{saket2017,saket2018,derakhti2018,seiffert2018, derakhti2020,touboul2021} have verified this result for a range of two-dimensional (2D) and three-dimensional (3D) wave packet types, water depths and wind forcing speeds. Furthermore, it has been shown \cite{derakhti2020,na2020} that the normalized rate of change of  $B$ as it passes through $B_{th}$ accurately predicts the breaking strength parameter $b$ \cite{phillips1985}, which quantifies the energy dissipated through breaking \cite[e.g.][]{drazen2008, Deike2015, sutherland2015}. Despite these significant findings, calculation of $B$ requires the measurement of both the local wave surface particle velocity and the crest speed, for which accurate measurements can be non-trivial.

\begin{figure}
    \centering
    \includegraphics[width=\textwidth]{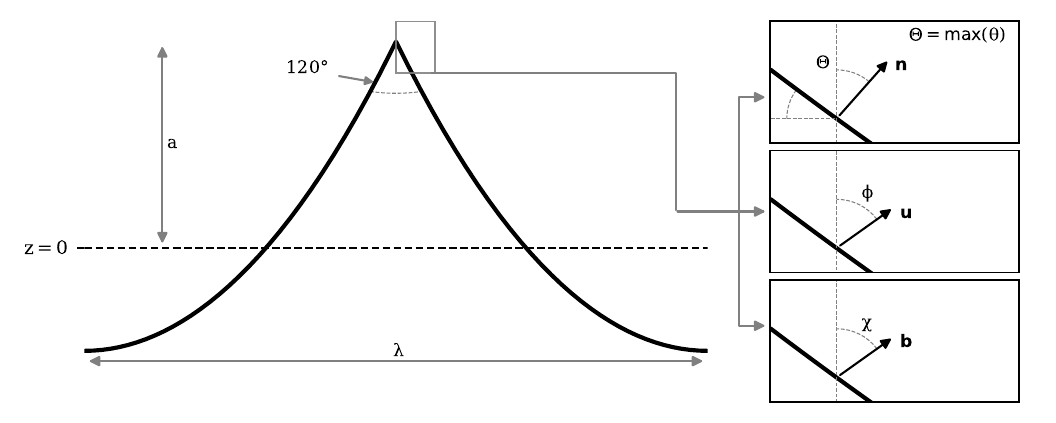}
    \caption{Schematic of a Stokes wave with limiting steepness $ak = 0.443$, where $a$ is the crest amplitude and $k = 2\pi / \lambda$ the wavenumber. The limiting steepness results in a crest angle of  $120 \degree$; this equates to a maximum local interface slope $\eta_x = \partial \eta / \partial x = -0.5774$ or \thetamax{} $=\max [\theta(x)]= 30 \degree$ (top inset). The kinematics of the wave motion (Section \ref{s:geometrics-linking}) can be further defined by the interface normal vector $\mathbf{n}$, which is at an angle \thetamax{} to the vertical, the surface particle velocity vector $\mathbf{u}$ with angle $\phi$ to the vertical (middle inset), and the boundary velocity vector $\mathbf{b}$ with angle $\chi$ (bottom inset).}
    \label{fig:crest-schematic}
\end{figure}

In contrast, geometric breaking inception parameters are based solely on the shape of the wave and in principle can be the easiest methods to implement accurately. Geometric parameters have a long history, beginning with the seminal work of \citet{stokes1880}. In an extension to his method for solving the gravity wave dynamic free-surface boundary condition \cite{stokes1847}, he found that for 2D, periodic waves in deep water and in the absence of surface tension, a limiting steepness $ak = 0.443$ exists, where $a$ is the wave amplitude and $k$ the wavenumber. This equates to a crest tip angle of 120$\degree$ and a maximum local interface slope \thetamax{} $=\max [\theta(x)]= 30 \degree$ relative to the horizontal (Fig. \ref{fig:crest-schematic}).

The relative simplicity of the Stokes model has led to its wide application for understanding linear wave dynamics. However, real waves are rarely periodic, stationary or linear and have been observed to break at both lower and higher $ak$ \cite{perlin2013}, with this variability attributed to a number of influences including wind \cite{phillips1974, kharif2008, touboul2006}, currents \cite{vreica2022} and the  bandwidth of the wave spectrum \cite{mcallister2023, pizzo2019}. Consequently, other authors have explored alternative measures of wave steepness as a breaking onset threshold. One such parameter is the theoretical maximum steepness $S = \Sigma k_n a_n$ \cite{drazen2008}, which sums the wavenumber $k_n$ and amplitude $a_n$ over $n$ Fourier wave components. However, the critical value of $S$ identified with these approaches varies considerably between studies \cite{perlin2013}. While much of this variability can be explained by the introduction of the wave packet bandwidth as a second dependent parameter, the method in which both $S$ and the bandwidth are defined can significantly impact results and a universal inception threshold based on the wave geometry is yet to be identified. 

More recently, interest has turned towards local measures of the slope of the interface $\eta$, defined as $\eta_x = \partial \eta / \partial x$ (or $\nabla \eta = \left[\partial \eta / \partial x, \partial \eta / \partial y\right]$ in 3D). Wave breaking onset is typically defined as the instant at which the maximum local interface angle \thetamax{} $=\max [\theta(x)]= 90 \degree$, that is, when $\eta_x \to -\infty$. This makes the local interface angle a logical choice for a breaking inception diagnostic parameter. In a numerical simulation study, \citet{mcallister2023} found that for 2D inviscid wave fields, a threshold value for the modulus of the minimum local interface slope 
\begin{equation}\label{eq:eta-x}
    \vert \min (\eta_x) \vert = 1/ \tan(\pi/3) \approx 0.5774
\end{equation}
separated breaking and non-breaking waves, which is equivalent to a threshold value for the maximum local interface angle
\begin{equation}
    \Theta_\text{th} = 30 \degree.
\end{equation}
Remarkably, this threshold angle is equivalent to the limiting steepness derived by \citet{stokes1880} and, for 2D waves, appears to be consistent regardless of the bandwidth of the wave group. However, recent laboratory results \cite{mcallister2024} indicate that for 3D wave fields the threshold value increases with increasing directional spread to approach the theoretical limit of $\Theta_\text{th} = 45 \degree$ for inviscid standing waves \cite{masden2006}. 

While these results offer promise of a simple yet effective breaking inception diagnostic, the localized nature of \thetamax{} means that its value can be highly dependent upon the measurement resolution. The 2D numerical experiments \cite{mcallister2023} employed a boundary element method \cite{dold1985} that, while computationally efficient, featured relatively coarse particle spacing at the interface equating to approximately 105 particles per wavelength for the highest resolution cases. Similarly, the 3D laboratory experiments \cite{mcallister2024} utilized an array of wave gauges at \SI{0.1}{\meter} spacing (equivalent to approximately 1/27 of the peak wavelength for forcing frequency \SI{0.75}{\hertz}), which the authors assessed underestimated the true value of \thetamax{} by approximately \ang{6}. 

A further limitation of the numerical study was the absence of surface tension, which acts as an additional force along the curved air-water interface. The importance of the surface tension force relative to gravity is quantified with the non-dimensional Bond number \cite[][Ch. 4]{kundu2011}. The geometry of the scenario is approximated with a length scale representing the radius of curvature $R_c$ over which the surface tension force is operating. In gravity wave studies, $R_c$ is typically approximated by the wavelength $\lambda$, such that \cite{Schwartz1982, Deike2015}
\begin{equation}\label{eq:bond-number}
    \text{Bo} = \frac{\rho_w g}{\sigma k^2},
\end{equation}
with water density $\rho_w$, gravity $g$, surface tension coefficient $\sigma$ and wavenumber $k = 2 \pi / \lambda$. For quasi-linear Stokes-like waves, $R_c$ is near-uniform across the wave interface and is well approximated by $\lambda$. In these cases, surface tension effects are generally considered significant for $\text{Bo} < O(10^3)$ (i.e. $\lambda < O(1)$ \si{\meter}). However, as waves become steeper and more nonlinear the crest tip becomes sharper and $\lambda$ is no longer representative of $R_c$ at the crest tip. As a consequence, surface tension has been shown to modify wave geometry even for large \bond{} \cite{Schwartz1982, Perlin2000}. For example, the limiting steepness separating breaking and non-breaking waves with $\text{Bo} = O(10^3)$ is larger than that of pure gravity waves \cite{Debiane1996}. 

A feature of steep gravity waves with weak surface tension is the formation of a bulge on the forward face of the crest that does not necessarily lead to wave breaking \cite{longuet1996, longuet1997}. While the majority of gravity wave studies focus on those waves that are breaking, there are a number of examples of highly non-linear non-breaking waves in previous studies \cite{Deike2015, boettger2023}. Details of the local geometry of these waves, specifically the value of \thetamax{} for maximally steep non-breaking cases, are unknown. 

The energetic, kinematic and geometric diagnostics outlined above represent distinctly different approaches to characterize breaking inception, yet in essence they each offer alternative views of the same physical process. This raises the question of whether these diagnostics together present a holistic view of breaking inception. \citet{mcallister2023} found that geometric inception preceded kinematic inception by approximately 0.3 wave periods in their simulations. This corresponds with the timing of energetic inception \citep{boettger2023, boettger2024}, suggesting possible links between the two. However, the effect of surface tension on the timing and magnitude \thetathresh{} is yet to be explored; this critical aspect should be resolved before \thetathresh{} could be used to understand wave breaking in realistic air-water gravity waves. 

The aim of this paper is firstly to determine how the threshold value \thetathresh{} for 2D waves is modified by the influence of surface tension, and secondly to explore the relationship between the energetic, kinematic and geometric inception diagnostics. Our approach is to utilize an ensemble of high-resolution direct numerical simulations (DNS) of mechanically generated breaking and non-breaking waves with varying wind forcing speeds, water depths and wave packet sizes, including the influence of surface tension. For each wave in this ensemble, we track the temporal evolution of \thetamax{} and quantify the impact of surface tension on these values using a local formulation of \bond{}. We further demonstrate the link between \thetamax{} and the kinematic parameter $B$ through a simple mathematical model and show that they track equivalent properties of the wave evolution. 

\section{Methodology}\label{s:methodology}

To investigate the geometric properties of wave breaking inception, we utilize an ensemble of non-breaking and breaking waves \cite{boettger2024data} simulated in a 2D Numerical Wave Tank (NWT). While limiting the study to 2D waves precludes an examination of the effects of wave directionality, the computational saving enables us to examine a wide range of other relevant wave parameters. The ensemble includes waves with varying wave packet size, water depth and wind forcing speeds (Table \ref{tab:experiments}) and consists of simulations previously undertaken to investigate energetic wave breaking inception \cite{boettger2023, boettger2024}. Here, we provide an overview of the NWT and the wave ensemble, and refer the reader to \citet{boettger2023,boettger2024} for further details. 

The NWT is based upon the DNS Navier-Stokes solver Gerris \cite[][]{popinet2003, popinet2009}, which implements the volume-of-fluid method to simulate two-phase, incompressible flow, including the effects of surface tension and viscosity. The NWT setup is similar to a laboratory experiment in which a fan drives a steady wind flow and waves are mechanically generated by a paddle \cite[e.g.][]{banner1990, reul1999, peirson2008, grare2013}. The wind-generated boundary layer at the inflow boundary $U_b(z)$ is described in \citet{boettger2024}. It was derived by applying a uniform air flow with velocity $U_0$ to an initially quiescent NWT, and fitting the analytical boundary layer solution of \citet{lock1951}. The paddle forcing was applied as an additional velocity and pressure gradient boundary condition derived from wavemaker theory for a bottom-mounted, flexible flap paddle \cite{dean1991}. The paddle signal consisted of a chirped packet function \cite{song2002} with a chirp rate $C_{ch}=1.0112 \times 10^{-2}$ and a paddle frequency $\omega_p$. The NWT was configured in non-dimensional coordinates utilizing scaling parameters derived from a deep water gravity wave with wavelength $\lambda_p=$ \SI{1}{\metre}, linear phase speed $c_p=$ \SI{1.15}{\metre\per\second}, and period $T_p=$ \SI{1}{\second}, resulting in $\text{Bo} \sim 2500$. 

The numerical representation of surface tension has traditionally been challenging as the accurate depiction of the pressure jump across the fluid interface on a finite grid is not compatible with a continuous solution to the momentum equations. This can lead to the generation of spurious currents at or near the interface \cite{Popinet2018}. In Gerris, this problem is minimized by using an improved implementation of the continuum-surface-force approach \cite{Popinet2018}. This was validated against an exact analytical solution for small-amplitude capillary waves \cite{Prosperetti1981} and shown to result in significantly smaller errors than comparable numerical surface tension models \cite{popinet2009}.

Gerris uses a quadtree grid structure that enables efficient adaptive grid refinement \cite{popinet2003}. Each level of refinement divides the parent cell into four, resulting in a maximum resolution equivalent to a uniform mesh size of $2^n \times 2^n$, for $n$ refinement levels. Cost functions were employed to focus high resolution at the air-water interface and in regions of high vorticity such that computational costs are minimized while still resolving critical features of the flow. A series of convergence tests with grid refinement levels of up to $2^{11}$ (approximately 1740 cells per wavelength) showed that a refinement level of $2^{10}$ (870 cells per wavelength) was sufficient to resolve the energetic processes within the wave \cite{boettger2023}.

The wave ensemble \cite{boettger2024data} consists of 173 simulations from 6 experiments (Table \ref{tab:experiments}), in which the wind speed $U_0/c_p$, number of waves in the paddle signal $N$ and water depth $d/\lambda_p$ were varied. Within each simulation, the evolution of each individual wave is independently tracked and the properties of the wave recorded as a function of both space and time. High-frequency variability in the temporal evolution of these properties was removed with the application of a running mean of width $0.15T_p$ and the variability around this mean was quantified using a time series bootstrap method \cite{boettger2023}. The local properties were tracked at two locations. The first, $\mathbf{x_+} = [x_+(t), z_+(t)]$ is where the local particle speed $\Vert\mathbf{u}\Vert$ has its maximum value and moves with velocity $\mathbf{b}_+=d \mathbf{x_+} / d t$. The second, $\mathbf{x_\theta} = [x_\theta(t), z_\theta(t)]$ is where the local interface angle has its maximum value \thetamax{} (Fig. \ref{fig:crest-schematic}). The distance between $\mathbf{x_+}$ and $\mathbf{x_\theta}$ as a wave approaches breaking inception is small ($\vert  \mathbf{x_+} - \mathbf{x_\theta} \vert \approx 1 \times 10^{-2} \lambda_p$) and the local velocities at these locations are similar, however, we found that using the particle velocity at $\mathbf{x_+}$ was necessary to accurately capture the kinematic inception threshold \cite{boettger2023}.

The volume-of-fluid formulation tracks the evolving fluid phase using a scalar field $\mathcal{T}$ that takes values of 0 or 1 to represent cells containing purely air or water and intermediate values for cells containing a mixture of fluids. The air-water interface $\eta$ is  calculated as the $\mathcal{T}=0.5$ contour. The onset of wave breaking is typically defined as the instant at which the local interface slope exceeds the vertical, that is, \thetamax{} $= 90 \degree$. Because of the high spatial resolution of the simulations, there are some cases in which only a single segment of the interface exceeds this threshold before again relaxing, in which case it is difficult to definitively classify a wave as either non-breaking or breaking. Consequently, we define a crest as breaking if the slope of the interface contour exceeds the vertical by a horizontal distance $\delta x\ge 0.5 {dx}$ over a length $\delta z \ge {dx}$, where $dx$ is the numerical grid spacing. Furthermore, the instant of breaking onset was defined as the first time that these thresholds are exceeded. A total of 20 marginal cases in which $-0.5 {dx} \ge \delta x \ge 0.5 {dx}$ are discarded from the ensemble, and the remaining cases classified as breaking or non-breaking. Using this approach, the final ensemble consists of 143 breaking and 941 non-breaking waves. 

To enable comparison of the crest energetics across all waves in the ensemble, a crest reference location and time are set as $[x_0, t_0] = [x_+, t]$ at the instant of breaking onset for breaking crests and at the instant of maximum $\Vert\mathbf{u}\Vert$ at \hotspot{} for non-breaking crests. The evolution of the crest in space and time is then referenced to these parameters using the non-dimensional coordinates $x^* = (x - x_0) / \lambda_p$,  $z^* = z / \lambda_p$ and $t^* = (t - t_0) / T_p$. 

\begin{table}
  \begin{center}
\def~{\hphantom{0}}
  \begin{fancytblr}
  \SetCell[r=2]{c} Experiment & \SetCell[r=2]{c} $U_0 / c_p$ & \SetCell[r=2]{c} $N$ & \SetCell[r=2]{c} $d/\lambda_p$ & \SetCell[r=2]{c} $2^n$ & \SetCell[r=2]{c} Simulations & \SetCell[c=2]{c} Total waves \\
       &  &  &  & & & B & NB \\
    \hline
    W0N5D2 & 0 & 5 & 0.17 & 10 & 9 & 6 & 32 \\
    W0N5D5 & 0 & 5 & 0.5 & 10, 11 & 49 & 30 & 274 \\
    W0N9D5 & 0 & 9 & 0.5 & 10, 11 & 38 & 38 & 211 \\    
    W1N5D5 & 1 & 5 & 0.5 & 10, 11 & 20 & 13 & 134 \\
    W2N5D5 & 2 & 5 & 0.5 & 10, 11 & 27 & 20 & 148 \\
    W6N5D5 & 6 & 5 & 0.5 & 10, 11 & 30 & 36 & 142 \\
    \hline
    & & & & \SetCell[r=1]{r} Total: & 173 & 143 & 941 \\    
  \end{fancytblr}
 \caption{Summary of the numerical experiments \cite{boettger2024data} included in this study. Experiments W0N5D2, W0N5D5 and W0N9D5 span varying wave packet size $N$, water depth $d/\lambda_p$, and maximum grid refinement $2^n$ in the absence of wind forcing \cite{boettger2023}. Experiments W1N5D5, W2N5D5, W6N5D5 feature identical wave paddle and water depth configuration as W0N5D5 but introduce variable wind speed forcing $U_0 / c_p$ \cite{boettger2024}. The total number of simulations, breaking (B) and non-breaking (NB) waves in each experiment are also listed.}
  \label{tab:experiments} 
  \end{center}
\end{table}

\section{Geometric inception}

We examine the geometric inception threshold in terms of the maximum local interface angle \thetamax{} $=\max [\theta(x)]$. This is equivalent to the minimum local steepness $\left | \min \left( \eta_x \right ) \right|$  used by \citet{mcallister2023} (i.e.  \thetamax{} $= \tan^{-1} \left | \min \left( \eta_x \right ) \right|$) yet provides a more intuitive unit of measurement, noting that \thetamax{} $= 90 \degree$ corresponds to the generally adopted definition of breaking onset. 

We use three representative wave crests taken from experiment W0N5D5 (Table \ref{tab:experiments}) to illustrate the evolution of \thetamax{}. With the exception of the paddle forcing amplitude, the NWT settings used to generate each crest are identical. The NB0 crest ($A_p / \lambda_p = 0.03700$) is a non-breaking wave with small maximum steepness, whereas the NB1 crest ($A_p / \lambda_p = 0.04006$) is the steepest, most energetic non-breaking crest in the ensemble. Finally, the B1 crest ($A_p / \lambda_p = 0.04004$) is a weakly breaking example. Both the NB1 and B1 crests are also used by \citet{boettger2023} to illustrate the energetic inception parameter described therein. 
\begin{figure}
    \centering
    \includegraphics[width=\textwidth]{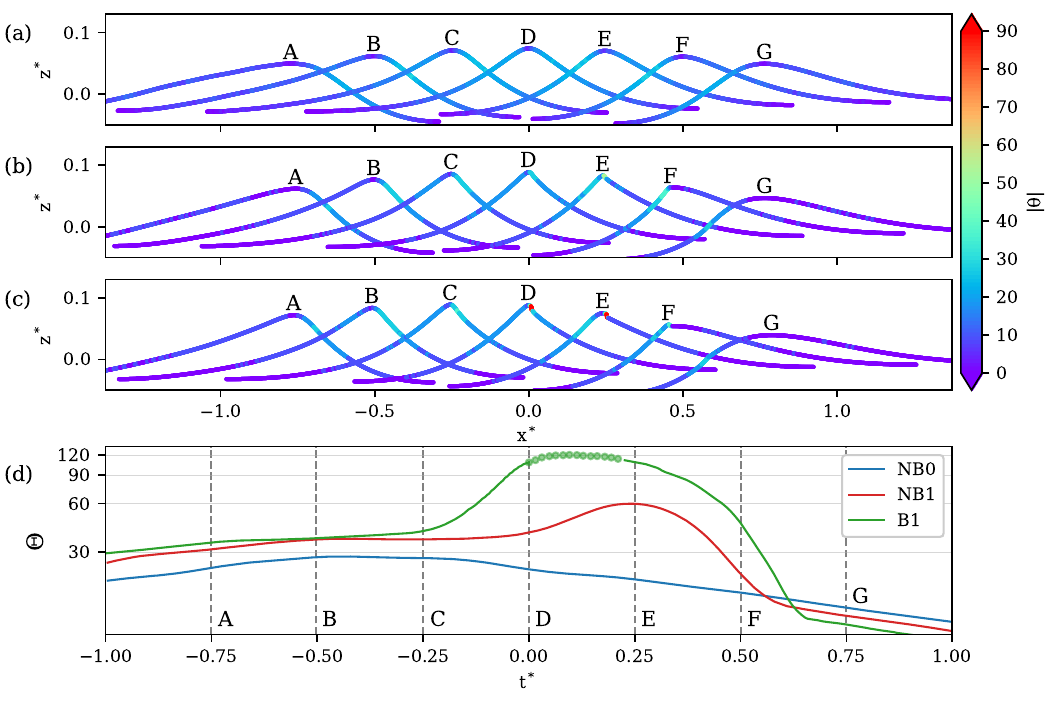}
    \caption{The evolution of the absolute value of the local interface angle $\vert \theta \vert$ for the (a) NB0 (non-breaking), (b) NB1 (near-breaking) and (c) B1 (breaking) representative wave crests as they progress through the growing and decaying phase. Snapshots are equally spaced at intervals of $0.25T_p$. In (d) the temporal evolution of \thetamax{} (using a power scale) for each wave is shown. The time of each snapshot (A-G) corresponds with the vertical dashed lines. Snapshot $D$ occurs at the time that $B$ at $\mathbf{x_+}$ has its maximum value for non-breaking waves, and the time of breaking onset for breaking waves. The total period of active breaking for the B1 wave crest is indicated by the dotted line.}
    \label{fig:theta-example-crests}
\end{figure}

The geometric evolution of these waves is shown in Fig. \ref{fig:theta-example-crests}. In panels a-c, snapshots of the wave interface are shown at 0.25$T_p$ intervals, with snapshot D corresponding to a time of $t^*=0$ and coinciding approximately with the instant of maximum wave amplitude. Each snapshot extends horizontally between the forward and rear trough minima, enabling the instantaneous geometric properties of each wave to be visually compared. On a broad scale, it can be seen that each wave undergoes a similar growth and decay cycle. Initially, the crest surges forward relative to the wave troughs, resulting in a highly asymmetric wave shape. Then, as the wave amplitude increases, the relative motion of the crest slows, so that by the time that the wave reaches its maximum amplitude (snapshot D), the wave is nearly symmetrical. The wave again becomes highly asymmetric as the crest leans backwards (snapshot G). A closer inspection does however, reveal some slight differences, which are most clearly evident at snapshot D. Here it can be seen that the additional energy provided by the larger paddle amplitude in the NB1 and B1 cases results in a larger crest amplitude, a shorter $\lambda$ and a correspondingly larger wave steepness $ak$. The crest also becomes noticeably sharper as the paddle amplitude increases, which can be quantified by the larger $\vert \theta \vert$ values in the vicinity of the crest. 

The largest $\vert \theta \vert$ values occur on the forward face of the crest tip and the evolution of the maximum value \thetamax{} for each wave (Fig. \ref{fig:theta-example-crests}d) illustrates their contrasting local geometry. The value of \thetamax{} for the NB0 wave increases smoothly to a maximum \thetamax{} $=29.5\degree$ (that is, just below the $30\degree$ threshold for waves without surface tension \cite{mcallister2023} ) at $t^*=-0.4$, before again decreasing. The evolution of \thetamax{} for the NB1 and B1 cases is strikingly similar to the NB0 crest for the periods up to $t^*=-0.25$ and after $t^*=0.6$, however, between these times the \thetamax{} values rapidly increase up to maximum values of  $59.9\degree$ and $120\degree$ respectively. The increased values of \thetamax{} in the NB1 wave is associated with the formation of a highly localized bulge on the forward face of the crest tip, consistent with the previous laboratory and numerical studies described in Section \ref{s:intro}. A similar bulge is also evident for the B1 crest, however, in this case it continues to grow and eventually overturns at breaking onset. 

\begin{figure}
\centering    
    \includegraphics[trim={1cm 0 0 0},clip,width=\textwidth]{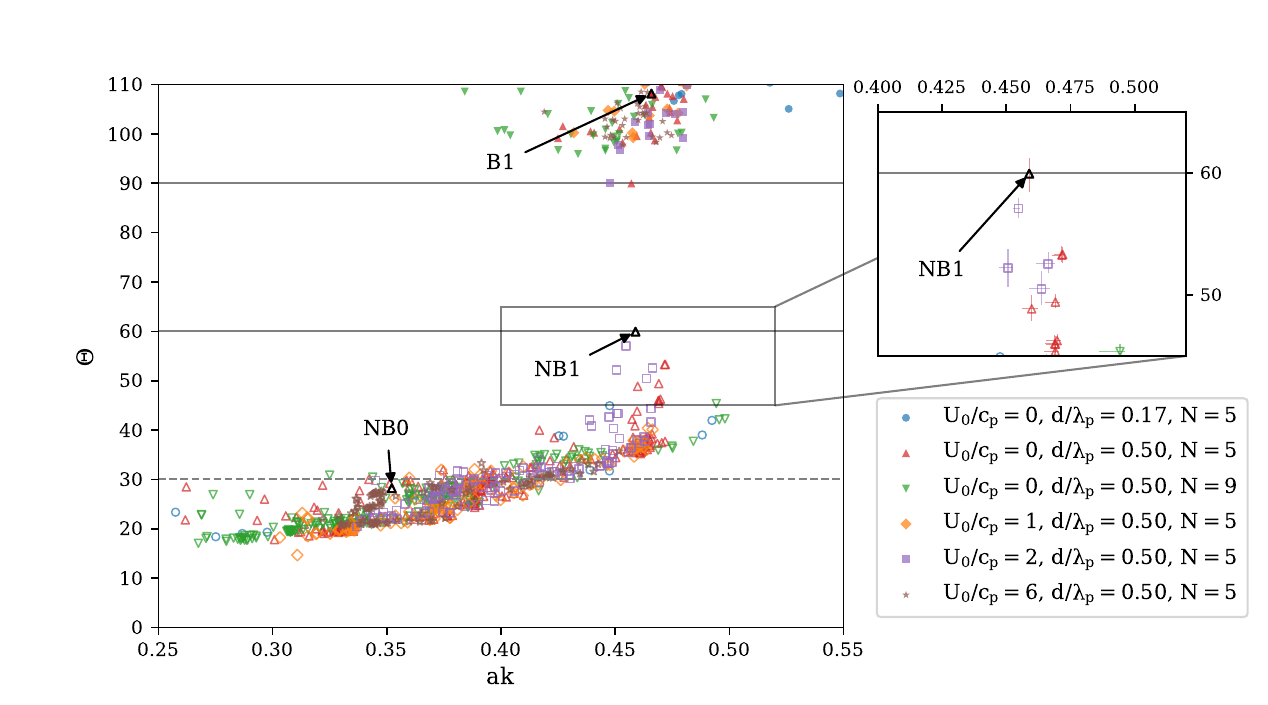}    
    \caption{The maximum value of \thetamax{} (non-breaking crests, hollow symbols) and the value at breaking onset (breaking crests, solid symbols) as a function of wave steepness $ak$. The \thetathresh{}$=30$ breaking inception threshold \cite{mcallister2023} for waves in the absence of surface tension is shown by the gray dashed line. Error bars are shown for the maximally steep non-breaking waves in the inset and represent 5 \% and 95 \% confidence intervals. For the maximally steep non-breaking wave NB1, \thetamax{} $=59.92 \pm [1.45, 1.25]\degree$.}
    \label{fig:theta-summary}
\end{figure}

The local geometry of the NB1 wave is not unique and a large number of non-breaking waves exceed the \thetathresh{}$=30 \degree$ threshold \cite{mcallister2023} to approach a limiting value of \thetamax{} $=60\degree$ (Fig. \ref{fig:theta-summary}). There is a clear separation between these maximally steep non-breaking crests and those that break (\thetamax{} $\ge 90 \degree$). The value of \thetamax{} does not show any significant dependence on wave packet size or wind speed for the range of parameters sampled. 

The maximum $ak$ of non-breaking waves has been shown to be dependent on the bandwidth of the wave spectrum \cite{mcallister2023, pizzo2019}, and, while \citet{mcallister2023} found that the \thetathresh{}$=30\degree$ inception threshold was consistent across a range of wave packet bandwidths, it is reasonable to investigate whether bandwidth has any impact on the above results. However, in terms of the global packet steepness $S = \Sigma k_n a_n$ \cite{drazen2008}, where $ak$ is summed over $n$ wave components, the chirped wave packets utilized here ($S \in [0.05, 0.09]$) are within the range of $S \in [0.02, 0.35]$ sampled by \citet{mcallister2023}, suggesting that these contrasting results cannot be attributed to the wave packet bandwidth. 

A more likely cause for these differences is the effect of surface tension, which is explicitly included in the current study but is absent in the simulations of \citet{mcallister2023}. The non-dimensional scaling used for the NWT is based on a deep water gravity wave with wavelength $\lambda_p=$ \SI{1}{\metre} (Section \ref{s:methodology}), resulting in $\text{Bo} \sim 2500$ using (\ref{eq:bond-number}) and the wavenumber derived from the paddle forcing $k_p = 2\pi / \lambda_p$, which suggests that gravity is the dominant restoring force over surface tension. However, the use of a fixed length scale $k_p$ underestimates the curvature of the crest tip for steep, nonlinear waves. A more precise formulation that accounts for the non-linear shape of the evolving wave crest is
\begin{equation}\label{eq:crest-bond-number}
    \text{Bo}_c = \frac{\rho_w g}{\sigma k_c^2},
\end{equation}
which utilizes the time-varying radius of curvature of the crest tip $R_c(t)$ to define the local crest wavenumber $k_c(t) = 2\pi / (4R_c)$. 

\begin{figure}
    \centering
    \includegraphics[width=\textwidth]{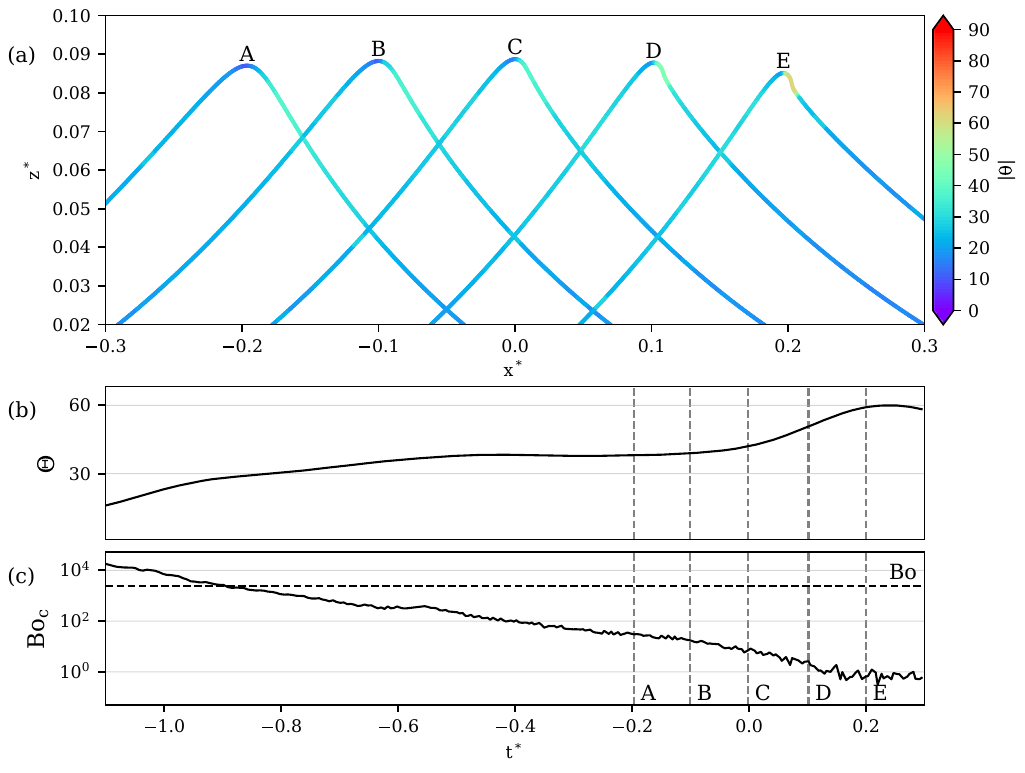}
    \caption{The evolution of the interface of the NB1 non-breaking crest. In (a), interface snapshots are shown at $0.1t/T_p$ intervals and colored by the absolute value of the local interface angle $\vert \theta \vert$. The time evolution of \thetamax{} is shown in (b), with the Bond number $\text{Bo}_c$ based on the time-varying crest tip curvature shown in (c). The Bond number \bond{} based on wave packet linear scaling is indicated by the black dashed line in (c).}
    \label{fig:theta-vs-bond-NB1}
\end{figure}

A closer examination of the NB1 non-breaking crest illustrates the evolution of the crest bulge and $\text{Bo}_c$ (Fig. \ref{fig:theta-vs-bond-NB1}). Over the final period of wave growth the crest steepens and the radius of curvature of the crest tip visibly reduces (Fig. \ref{fig:theta-vs-bond-NB1}a). The maximum angle \thetamax{} increases concurrently with increasing crest non-linearity and reaches a maximum value of $59.2 \degree$ at $t^* \sim 0.2$. The local Bond number $\text{Bo}_c$ decreases in line with the increasing magnitude of \thetamax{}, passing through $\text{Bo}$ as \thetamax{} exceeds $30 \degree$. The magnitude of $\text{Bo}_c$ continues to decrease to a minimum value $\text{Bo}_c \approx 1$, illustrating the increasing significance of surface tension on the local crest geometry. 

\begin{figure}
    \centering
    \includegraphics[width=\textwidth]{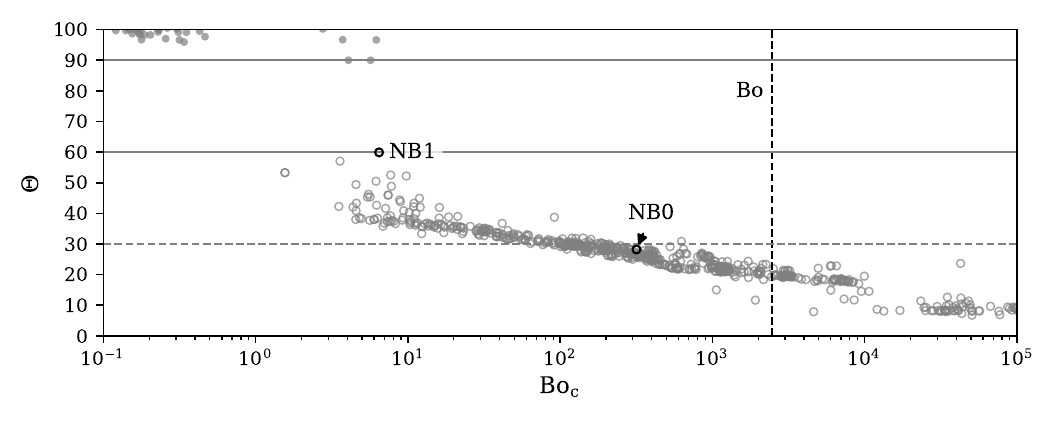}
    \caption{The maximum value of \thetamax{} (non-breaking crests, hollow symbols) and the value at breaking onset (breaking crests, solid symbols) as a function of the crest Bond number $\text{Bo}_c$ from (\ref{eq:crest-bond-number}). The Bond number $\text{Bo}$ (\ref{eq:bond-number}), formulated using the paddle forcing wavelength as the length scale, is indicated by the dashed line.}
    \label{fig:theta-vs-Bo}
\end{figure}

All of the non-breaking crests in the ensemble with \thetamax{} $> 30 \degree$ are associated with $\text{Bo}_c < 10^2$ (Fig. \ref{fig:theta-vs-Bo}). It is evident from these results that, for the gravity waves with $\lambda\approx1$ \si{\meter} simulated in this study, the presence of surface tension significantly influences the local crest shape and is likely the leading factor responsible for the \thetathresh{}$= 60 \degree$ threshold separating breaking and non-breaking crests, in contrast to the \thetathresh{}$= 30 \degree$ threshold reported by \citet{mcallister2023} in their experiments without surface tension.  

\section{Relation between geometric and kinematic inception}
\label{s:geometrics-linking}

Local diagnostic parameters for wave breaking inception, such as \thetamax{} and the energetic \cite{boettger2023, boettger2024} and kinematic \cite{barthelemy2018} diagnostics methods discussed in Section \ref{s:intro}, are all focused on the properties of the wave at the crest tip where the wave maximum energy and maximum interface slope coincide with the location of initial breaking onset. This raises the question of whether these complementary approaches are capturing different aspects of the same dynamical process. 
We explore an aspect of this question by considering the relation between the geometric and kinematic approaches.

The kinematic inception parameter (\ref{eq:B}) consists of an energy flux velocity $\mathbf{u}$ and the normalizing velocity $\mathbf{c}$. 
While the flux velocity is typically sampled at the location \hotspot{} near the crest tip that has the maximum value $\mathbf{u}_+$, the unsteady motion of the crest \cite{fedele2020} makes determining a representing crest velocity challenging. \citet{barthelemy2018} defined $\mathbf{c} = d \mathbf{x}_c /dt$ with the highest point of the crest tip $\mathbf{x}_c$, although alternate methods have been explored \cite{Craciunescu_2019, stansell2002}. \citet{boettger2023} showed that (\ref{eq:B}) can be formulated using the velocity $\mathbf{b}_+=d \mathbf{x_+} / d t$, such that
\begin{equation}\label{eq:B-plus}
    B = \frac{\| \mathbf{u}_+ \|}{\| \mathbf{b}_+ \|},
\end{equation}
with threshold value \Bth{}$=0.855 \pm 0.005$ as found by \citet{barthelemy2018}. For the waves in our ensemble, the difference between $\Vert\mathbf{c}\Vert$ and $\Vert \mathbf{b_+} \Vert$ for $B\approx B_\mathrm{th}$ is less than 1 \%. Similarly, we note that $\mathbf{x_+}$ is closely collocated with the position $\mathbf{x_\theta}$ of \thetamax{} ($\vert\mathbf{x_+} - \mathbf{x_\theta} \vert \approx 1 \times 10^{-2} \lambda_p$ for $B\approx B_\mathrm{th}$), such that it is reasonable to approximate $\mathbf{x_+} \approx \mathbf{x_\theta}$. Therefore, we can show that the 
vectors $\mathbf{u}_+$ and $\mathbf{b}_+$ are related to \thetamax{} through their normal components as
\begin{subequations}\label{eq:udotn}
\begin{gather}
    \mathbf{u}_+ \cdot \mathbf{n} = u \sin \Theta + w \cos \Theta, \\ 
     \mathbf{b}_+ \cdot \mathbf{n} = b_x \sin \Theta + b_z \cos \Theta. 
\end{gather}
\end{subequations}
Applying the trigonometric identity
\begin{equation}
    a \cos \Theta + b \sin \Theta = c \cos (\Theta +  \varphi)
\end{equation}
where $\varphi = \tan^{-1}\left(-{b}/{a} \right)$ and $c = \text{sgn}(a) \sqrt{a^2 + b^2}$, (\ref{eq:udotn}) can be written as
\begin{subequations}\label{eq:udotn-angle}
\begin{gather}
    \mathbf{u}_+ \cdot \mathbf{n} = \text{sgn}(w)\Vert \mathbf{u}_+ \Vert \cos (\Theta - \phi), \\
    \mathbf{b}_+ \cdot \mathbf{n} = \text{sgn}(b_z)\Vert \mathbf{b}_+ \Vert \cos (\Theta - \chi),
\end{gather}
\end{subequations}
where $\phi$ and $\chi$ are the angles relative to the vertical coordinate of $\mathbf{u}_+$ and $\mathbf{b}_+$ respectively (Fig. \ref{fig:crest-schematic}). The interpretation of (\ref{eq:udotn-angle}) is straightforward. If, for example, the velocity vector $\mathbf{u}_+$ is collinear with the unit normal vector $\mathbf{n}$, then $\Theta - \phi=0$ and  $\mathbf{u}_+ \cdot \mathbf{n} = \Vert \mathbf{u}_+ \Vert$. Conversely, any misalignment between the direction of $\mathbf{u}_+$ and $\mathbf{n}$ results in a corresponding decrease in the magnitude of $\mathbf{u}_+ \cdot \mathbf{n}$. 

For a material surface, such as the air-water interface,
\begin{equation}\label{eq:material-surface}
    \mathbf{u}_+ \cdot \mathbf{n} = \mathbf{b}_+ \cdot \mathbf{n}
\end{equation}
such that (\ref{eq:B-plus}) and (\ref{eq:udotn-angle}) can be combined to derive
\begin{equation}\label{eq:angle-ratio}
    B = \frac{\Vert \mathbf{u}_+ \Vert}{\Vert \mathbf{b}_+ \Vert} = \frac{\vert \cos (\Theta - \chi)\vert}{\vert\cos (\Theta - \phi)\vert},
\end{equation}
which shows that $B$ is equivalent to the ratio of the collinearity of the velocity vectors $\mathbf{u}_+$ and $\mathbf{b}_+$ with respect to the local interface angle \thetamax{}. 

\begin{figure}
\centering
    \includegraphics[trim={1cm 0 0 0},clip,width=\textwidth]{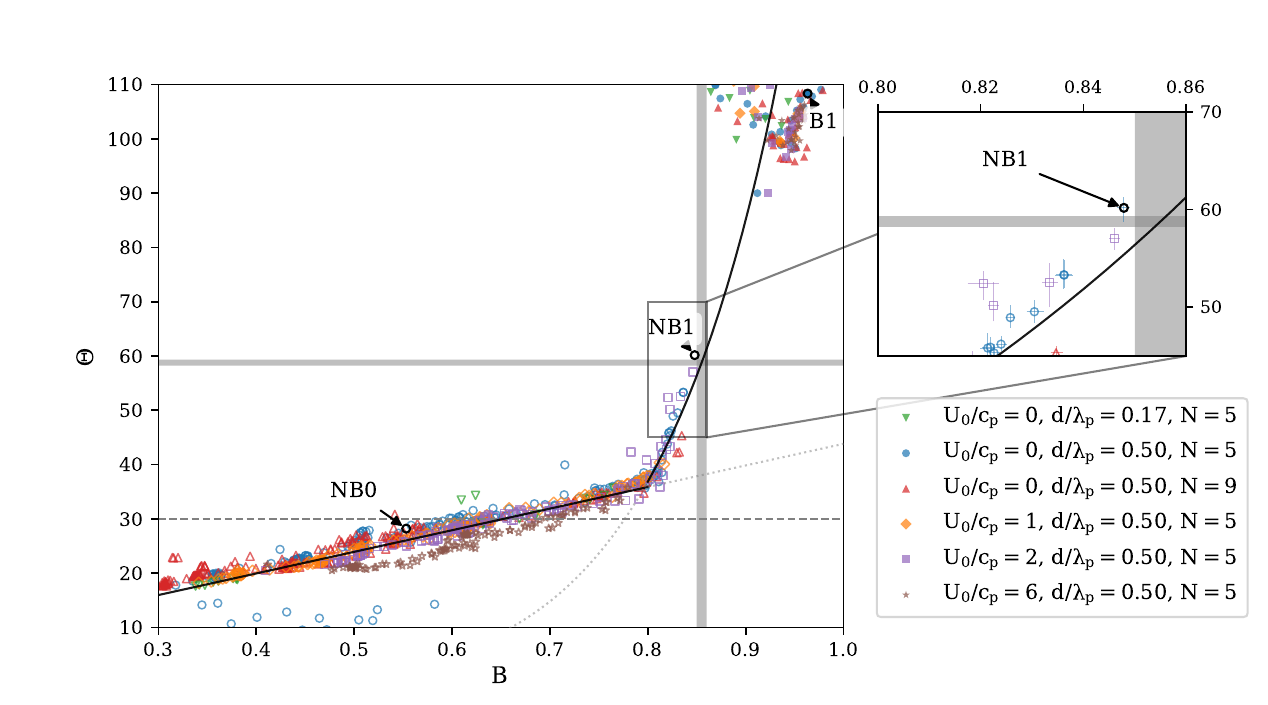}
    \caption{The maximum value of \thetamax{} (non-breaking crests, hollow symbols) and the value at breaking onset (breaking crests, solid symbols) as a function of $B$. The two-part fit function (\ref{eq:theta-fit}) is shown by the black line. The threshold value $B_{th} = 0.855 \pm 0.05$ \cite{barthelemy2018} and its equivalent from (\ref{eq:Bth-equiv-theta}) \thetathresh{}$ = 58.76 \pm 0.55 \degree$ are shown by the shaded gray regions, and the \thetathresh{}$=30\degree$ threshold for waves without surface tension \cite{mcallister2023} by the dashed line. Error bars on the data in the inset represent the 5 \% and 95 \% confidence intervals. For the maximally steep non-breaking wave NB1, $B= 0.848 \pm [-0.001, 0.001]$ and \thetamax{} $=59.92 \pm [1.45, 1.25]\degree$.}
    \label{fig:B-vs-theta-summary}
\end{figure}

The interaction between each of the terms in (\ref{eq:angle-ratio}) as a wave evolves is complex, and the resulting relationship between \thetamax{} and $B$ is non-linear (Fig. \ref{fig:B-vs-theta-summary}). Two distinct regimes are evident, for which \thetamax{} initially increases quasi-linearly as a function of $B$ before increasing more rapidly up to and beyond breaking inception. These two regimes are separated by the point $[B, \Theta] \approx [0.8, 37\degree]$, which coincides with the location at which $\text{Bo}_c = 10$ and a distinct bulge begins to form on the crest of steep non-breaking cases \mbox{(Figs. \ref{fig:theta-vs-bond-NB1}--\ref{fig:theta-vs-Bo})}. This relationship can be described empirically by (Fig. \ref{fig:B-vs-theta-summary}, black line)
\begin{equation}\label{eq:theta-fit}
    \Theta = 
    \begin{cases}
39.8 B + 4 & \qquad B \le 0.8 \\
\left(\frac{B}{B_{\textrm{th}}}\right)^{5.5} \arcsin B &  \qquad  B > 0.8.
    \end{cases}
\end{equation}
For $B =$ \Bth{}, (\ref{eq:theta-fit}) simplifies to
\begin{equation}\label{eq:theta-fit-Bth}
    \Theta(B_\text{th}) = \arcsin(B),
\end{equation}
which for \Bth{} $= 0.855 \pm 0.005$ corresponds to a threshold value for the maximum local interface angle of
\begin{equation}\label{eq:Bth-equiv-theta}
    \Theta_\text{th} = 58.76 \pm 0.55 \degree,
\end{equation}
shown by the horizontal shaded region in Fig. \ref{fig:B-vs-theta-summary}. 
The maximally steep NB1 case is seen to lie at the intersection of \Bth{} and \thetathresh{}, with \mbox{$B= 0.848 \pm [-0.001, 0.001]$} and \mbox{\thetamax{} $=59.92 \pm [1.45, 1.25]\degree$}. 

\begin{figure}
    \centering
    \includegraphics[width=\textwidth]{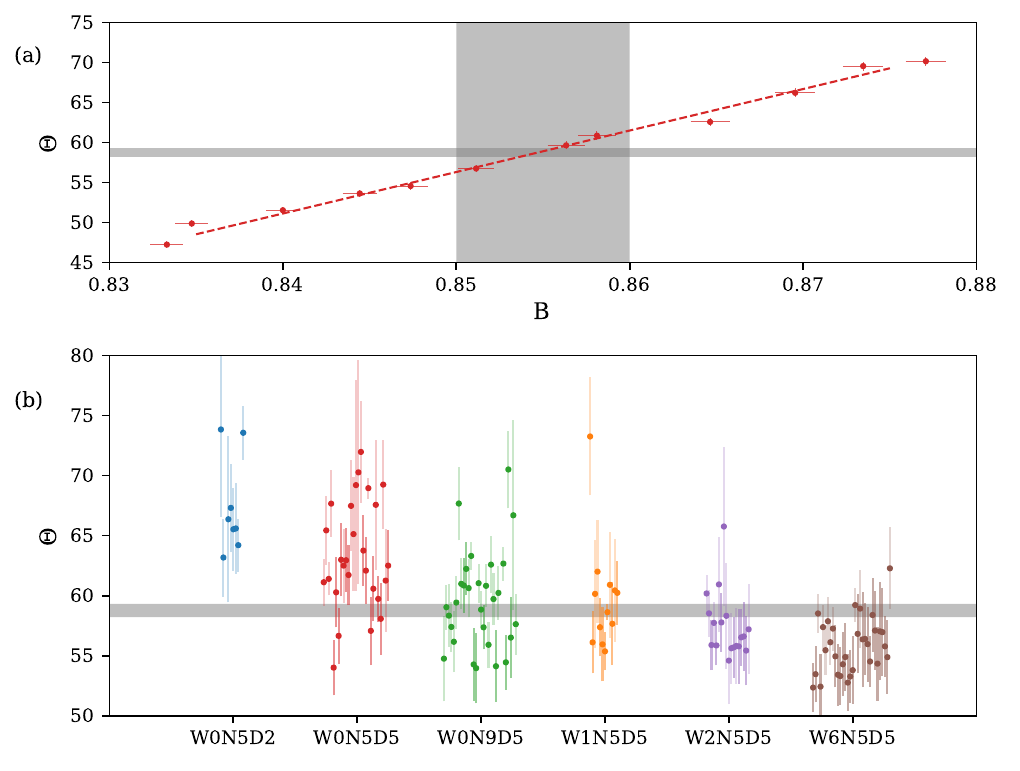}
    \caption{The magnitude of \thetamax{} as $B$ passes through \Bth{} for all breaking crests in our ensemble. In (a), the relation (\ref{eq:theta-fit-Bth}) is fit to the data for the B1 crest using a linear orthogonal distance regression over the interval $B \in B_\mathrm{th} \pm 0.02$ using the standard deviation shown by the error bars. The value of \thetamax{} at $B = 0.855 \pm 0.005$ is shown for all crests in (b), with data grouped and colored by the experiment number (Table \ref{tab:experiments}). The gray shaded regions correspond to $B_\mathrm{th} = 0.855 \pm 0.005$  and $\Theta_\text{th} = 58.76 \pm 0.55 \degree$.}
    \label{fig:theta-regression-B}
\end{figure}

The relation (\ref{eq:theta-fit-Bth}) also suggests that, for waves that go on to break, \thetamax{} and $B$ pass through their respective threshold values simultaneously. To examine this aspect, we calculate the value of \thetamax{} as $B \rightarrow B_\mathrm{th}$ using a linear orthogonal distance regression \cite{Boggs1990} over the interval $B \in B_\textrm{th} \pm 0.02$, with points weighted by the standard deviation of the bootstrap fit (Section \ref{s:methodology}) of both \thetamax{} and $B$. 
This process is shown for the representative B1 crest in Figure \ref{fig:theta-regression-B}a, from which the range of values corresponding to $B = 0.855 \pm 0.005$ is found to be \thetamax{} $=58.97 \pm 1.10 \degree$.  
The results for all breaking waves in the ensemble are shown in Fig. \ref{fig:theta-regression-B}b, for which the marker and error bars represent the value of \thetamax{} for $B = 0.855$ and the range $B = [0.85, 0.86]$ respectively.  In most cases these values are clustered around the shaded \thetathresh{} range, although there are some outliers. The outlying cases are characterized by large error bars (up to $\pm 9\degree$), which indicates that they are weakly breaking waves where $B$ slowly passes through and only just exceeds \Bth{}, making an accurate estimation of \thetamax{} more challenging. 

Other trends are also evident from the ensemble of results in Fig. \ref{fig:theta-regression-B}b. The waves from the intermediate water depth experiment W0N5D2 (blue markers) are all biased high, with a median \thetamax{} value of $65 \degree$ for $B = 0.855$. However, the number of breaking cases in this experiment is small and further experiments are necessary to determine whether the evolution of \thetamax{} is sensitive to water depth. 

\section{Discussion and conclusions}

In this paper, we have used an ensemble of surface gravity waves generated within a numerical wave tank to investigate the existence of a threshold value for the maximum local interface angle \thetamax{} that separates 2D breaking and non-breaking waves. Our work builds upon the recent efforts of \citet{mcallister2023}, who reported a threshold value of \thetathresh{}$=30 \degree$ \mbox{($\left | \min \left( \eta_x \right ) \right| = 0.5774$)} for 2D waves in the absence of surface tension. Here, our ensemble of simulations specifically includes surface tension effects and the resolution of the simulations (up to 1740 cells per wavelength) allows us to accurately determine the local geometric wave properties. Furthermore, the ensemble encompasses a range of wave packet size, water depth and wind speed forcing, enabling us to assess the sensitivity of  \thetamax{} to these parameters. 

Our results show that, in contrast to simulations without surface tension, non-breaking waves are limited to a maximum local interface angle \thetathresh{}$\approx 60 \degree$. For the steepest non-breaking waves, this manifests as a bulge on the forward face of the crest tip that increases the local interface angle relative to an inviscid wave. All waves that exceed this threshold value within our ensemble then proceed to break, regardless of the wave packet size, water depth or wind forcing speed. We demonstrate the role of surface tension in modifying the local wave geometry by tracking the wave crest Bond number $\text{Bo}_c$, which uses the minimum local radius of curvature $R_c$ as the length scale, and compare this to the Bond number \bond{} formulated with the wavenumber $k$. This shows that when \thetamax{} $=30 \degree$, $\text{Bo}_c \approx \text{Bo}$ and surface tension has an insignificant effect on the wave geometry. However, as \thetamax{} $\rightarrow60 \degree$, $\text{Bo}_c \rightarrow 1$ and surface tension is the dominant restoring force over gravity in determining the near-crest geometry.  

Wave breaking inception thresholds are typically founded upon dynamical, kinematic or geometric approaches \cite{perlin2013}. Dynamical approaches \cite{boettger2023, boettger2024} are necessary to understand the underlying physical processes leading to wave breaking. However, they can be difficult to implement outside of numerical simulations or laboratory experiments. In contrast, kinematic and geometric approaches represent a proxy of the underlying physics yet are simpler to measure. In this light, it is plausible that each approach provides a complementary view of the same phenomenon. 

This motivated our investigation of the relation between the geometric parameter \thetamax{} and the kinematic parameter $B$ \cite{barthelemy2018}. We show that for 2D wave crests, $B$ is equivalent to the ratio of the collinearity of the velocity vectors $\mathbf{u}$ and $\mathbf{b}$ with respect to \thetamax{}. As the crest approaches breaking inception and the directions of both $\mathbf{u}$ and $\mathbf{b}$ at the crest tip location \hotspot{} are nearly horizontal, the relationship further simplifies to an exact equivalence between $B$ and \thetamax{}. The geometric threshold corresponding to \Bth{} $=0.855 \pm 0.005$ is \thetathresh{}$= 58.76 \pm 0.55 \degree$, which we verify with our wave ensemble. This result suggests that the physical meaning of the kinematic and geometric inception thresholds may be related to the increasing misalignment of the $\mathbf{u}$ and $\mathbf{b}$ vectors relative to the interface angle $\theta$. 

While the threshold value of the kinematic inception parameter has been shown to be robust for a range of 2D and 3D wave packets in varying water depths and wind forcing speeds, the applicability of the geometric inception threshold to any arbitrary wave field is yet to be fully explored. Notwithstanding the limited spatial resolution obtainable in their laboratory experiments, \textcite{mcallister2024} demonstrated a consistent relationship between the magnitude of \thetathresh{} and the directional spread of 3D wave fields. In our ensemble of 2D numerical simulations, the magnitude of  \thetathresh{} was constant across all wave packet types and wind forcing speeds, however, the ensemble contained only a limited set of intermediate-depth waves (experiment W0N5D2), which was insufficient to definitively determine the sensitivity of \thetathresh{} to water depth. Furthermore, the parameter space explored in this paper did not examine the impact of the carrier wavelength on \thetathresh{} and the results are based on a dimensional scaling of $\lambda=O(1)$ \si{\meter}.  With consideration of the global Bond number (\ref{eq:bond-number}), it is possible that waves with a shorter or longer wavelength could exhibit a different threshold value. On a similar note, it is unclear how the value of \thetathresh{} would change as the surface tension coefficient $\sigma\rightarrow0$; would there be a smooth transition from the \thetathresh{}$=60\degree$ reported here to the \thetathresh{}$=30\degree$ result of \citet{mcallister2023} for $\sigma=0$, or a step change? Further studies examining these aspect are required before the universality of the \thetamax{} threshold is fully resolved.

\begin{acknowledgements}
This research was supported by the Commonwealth of Australia as represented by the Defence Science and Technology Group of the Department of Defence. Computational resources were provided through the National Computational Merit Allocation Scheme of the Australian Government's National Collaborative Research Infrastructure Strategy, as well as the University of New South Wales (UNSW) Resource Allocation Scheme. MLB acknowledges Australian Research Council support (grants DP120101701 and DP210102561) which motivated the present paper and its foundational contributions towards a more robust understanding of breaking water waves. 
\end{acknowledgements}

\bibliography{reference-library}

\end{document}